\documentclass[conference]{IEEEtran}
\usepackage{graphicx}
\usepackage{amsmath}
\usepackage{subcaption}
\usepackage{listings}
\usepackage{xcolor}
\usepackage{orcidlink}
\usepackage{hyperref}
\usepackage{algorithm}
\usepackage{algorithmic}
\usepackage{amssymb}

\begin{document}

\title{AEGIS: Adversarial Entropy-Guided Immune System\\
\Large Thermodynamic State Space Models for Zero-Day Network Evasion Detection}

\author{
\IEEEauthorblockN{Vickson Ferrel~\orcidlink{0009-0009-0155-0913}}
\IEEEauthorblockA{
\textit{Faculty of Computer Science \& Information Technology} \\
\textit{Universiti Malaysia Sarawak (UNIMAS)}, Kota Samarahan, Malaysia \\
\textit{Founder \& Lead Architect, Vixero Technology Enterprise}, Kuching, Sarawak, Malaysia \\
106641@siswa.unimas.my $|$ vickson@vixdev.cloud}
}

\maketitle

\begin{abstract}
As TLS 1.3 encryption increasingly limits the effectiveness of traditional Deep Packet Inspection (DPI), the security community has pivoted to Euclidean Transformer-based classifiers (e.g., ET-BERT) for encrypted traffic analysis. However, these content-reading models remain structurally vulnerable to byte-level adversarial morphing. Recent white-box pre-padding attacks have reduced ET-BERT accuracy to 25.68\%, while cryptographic mimicry protocols like VLESS Reality routinely bypass certificate-based detection. To address this fundamental limitation, we introduce AEGIS: an Adversarial Entropy-Guided Immune System powered by a Thermodynamic Variance-Guided Hyperbolic Liquid State Space Model (TVD-HL-SSM). AEGIS does not compete in the Euclidean payload-reading domain; rather, it establishes a distinct paradigm by discarding payload bytes entirely in favor of 6-dimensional continuous-time flow physics. We project these temporal features into a non-Euclidean Poincar\'e manifold, naturally accommodating the exponential branching of botnet topologies while mitigating gradient explosion. Concurrently, Liquid Time-Constants measure microsecond Inter-Arrival Time (IAT) decay, and a Thermodynamic Variance Detector calculates sequence-wide Shannon Entropy to expose the mathematical anomalies inherent to automated Command and Control (C2) tunnels. By coupling a pure C++ eBPF network ingress Harvester with a zero-copy Shared Memory (IPC) bridge, AEGIS bypasses the Python Global Interpreter Lock (GIL). Leveraging a linear-time $\mathcal{O}(N)$ Mamba-3 core compiled via TileLang, AEGIS processes 64,000-packet parallel swarms at line-rate without the memory bottlenecks characteristic of Transformers. Evaluated on a 400GB, 4-tier adversarial corpus—spanning trans-Pacific backbone traffic, IoT botnet captures, SHA256-verified zero-days, and proprietary VLESS Reality tunnels—AEGIS achieves a 0.9952 F1-score and a 99.50\% True Positive Rate. Operating at a Swarm Batch Size of 64, the architecture establishes a theoretical computational ceiling of 40 Million Packets Per Second (Mpps) with a 1.6~ms aggregate inference latency on a consumer RTX 4090. AEGIS establishes a new state-of-the-art for non-Euclidean, physics-based adversarial network defense, while identifying the theoretical detection limits of flow-based physics against true human-mimicry protocols.
\end{abstract}

\begin{IEEEkeywords}
Network Security, Encrypted Traffic Analysis, State Space Models, Hyperbolic Geometry, Deep Packet Inspection, Adversarial Evasion.
\end{IEEEkeywords}

\section{Introduction}
\label{sec:introduction}

The pervasive adoption of TLS 1.3 \cite{rfc8446} has severely limited the capabilities of traditional Deep Packet Inspection (DPI). By encrypting payload data and the Server Name Indication (SNI), modern cryptographic protocols reduce the efficacy of legacy firewalls that rely on signature matching and plaintext analysis. In response, the cybersecurity community has increasingly adopted Deep Learning (DL) architectures to classify encrypted traffic. Specifically, Euclidean Transformer-based models---such as ET-BERT \cite{lin2022etbert}---have emerged as a standard approach. These models leverage natural language processing (NLP) techniques to interpret packet byte sequences, frequently achieving $>$99\% accuracy on static, non-adversarial benchmarks \cite{yan2025certta}.

However, recent literature indicates a structural vulnerability within this paradigm: content-reading classifiers are inherently sensitive when subjected to adversarial morphing. Because models like ET-BERT prioritize the sequential order of payload bytes (often anchoring to the initial 32 bytes of a packet), attackers can effectively bypass detection through sequence manipulation. Jing et al. \cite{jing2025advtraffic} demonstrated that adversarial pre-padding techniques---injecting stochastic byte noise at the beginning of packets---can degrade ET-BERT’s classification accuracy to 25.68\%. Furthermore, cryptographic mimicry tools, such as VLESS Reality \cite{vless2023}, dynamically spoof legitimate TLS 1.3 WebSockets, reducing the reliability of certificate-based detection entirely.

Compounding this limitation is the industry's reliance on intent-based classification, where systems attempt to distinguish between benign encrypted tunnels (e.g., commercial VPNs) and malicious obfuscation. We argue that relying on this distinction presents a critical security gap. Adversarial morphing frameworks, such as AMOI \cite{ferrel2026amoi}, have demonstrated the ability to synthetically generate padding that anchors malicious Command and Control (C2) tunnels to the volumetric center (the ``Fat Middle'') of benign distributions. Consequently, distinguishing between benign and malicious obfuscation becomes highly challenging when the adversary actively controls the morphing algorithm. To maintain perimeter defense, network systems must adopt a stringent zero-trust classification paradigm: continuous-flow proxies, XTLS Reality tunnels, and encrypted obfuscation tools must be classified as anomalies, terminating unauthorized tunnels regardless of their simulated intent.

Moving beyond payload bytes to flow-level physics presents distinct challenges. Prior work on the AMOI framework revealed the ``Jitter Anomaly'' \cite{ferrel2026amoi}, demonstrating that dynamic morphing driven by stateless surrogates inadvertently triggers temporal alerts in stateful firewalls. Furthermore, advanced white-box adversarial attacks, such as Ayaka AH-MSI \cite{ferrel2026amoi}, have achieved evasion rates exceeding 99.5\% against standard continuous-time sequence models by effectively mimicking temporal distributions via Manifold Shattering.

To address these vulnerabilities, this paper introduces AEGIS: an Adversarial Entropy-Guided Immune System powered by a Thermodynamic Variance-Guided Hyperbolic Liquid State Space Model (TVD-HL-SSM). AEGIS operates in a distinct mathematical classification genre from Euclidean sequence models. Instead of processing susceptible payload bytes, AEGIS operates strictly on 6-dimensional flow physics. By projecting temporal features into a non-Euclidean Poincar\'e manifold \cite{nickel2017poincare}, AEGIS mitigates gradient explosion while accommodating the exponential branching of botnet topologies. Concurrently, Liquid Time-Constants \cite{hasani2021liquid} measure microsecond Inter-Arrival Time (IAT) decay, and a Thermodynamic Variance Detector calculates sequence-wide Shannon Entropy to identify the structural variance inherent to automated C2 tunnels, rendering adversarial pre-padding \cite{jing2025advtraffic} largely ineffective.

The main contributions of this study are summarized as follows:
\begin{itemize}
    \item \textbf{The TVD-HL-SSM Architecture:} We propose a neural architecture fusing Hyperbolic Poincar\'e embeddings \cite{nickel2017poincare}, Liquid Time-Constants \cite{hasani2021liquid}, and a hardware-aware Mamba-3 core \cite{lahoti2026mamba3}. Compiled via TileLang JIT, this configuration allows line-rate processing of 1,000-packet causal windows without the GPU memory constraints typical of $\mathcal{O}(N^2)$ Transformers.
    \item \textbf{Thermodynamic Entropy Detection:} To counter white-box manifold shattering, we introduce a Thermodynamic Variance Detector that calculates the Shannon Entropy of hidden states, identifying the unnatural variance of automated evasion tunnels.
    \item \textbf{Zero-Trust Adversarial Corpus:} We construct a 400GB, 4-tier adversarial corpus encompassing trans-Pacific backbone traffic, IoT botnets, zero-day rootkits, and proprietary VLESS Reality captures to evaluate cryptographic obfuscation under strict zero-tolerance parameters.
    \item \textbf{Performance Validation under Adversarial Conditions:} Evaluated on an RTX 4090 via bare-metal CUDA Graph optimization, AEGIS achieves an F1-score of 0.9952 and a 99.50\% True Positive Rate against zero-day evasion at a sub-millisecond inference latency of 262.27~$\mu$s, establishing a state-of-the-art baseline for continuous-time thermodynamic traffic classification.
\end{itemize}

\section{Background and Threat Model}
\label{sec:background}

To contextualize the architectural design of AEGIS, it is necessary to examine the evolution of encrypted traffic classification and formally define the adversarial capabilities that modern defense systems must withstand. 

\subsection{Limitations of Current Deep Learning Paradigms}
Historically, Deep Packet Inspection (DPI) relied on static signatures and heuristic rule-sets. As TLS 1.3 \cite{rfc8446} standardized payload and SNI encryption, research shifted toward Deep Learning (DL) approaches. Current state-of-the-art architectures largely utilize Euclidean Transformer models (e.g., ET-BERT \cite{lin2022etbert}) or Convolutional Networks (e.g., NetConv). These models process network packets analogously to natural language tokens, extracting latent representations from byte sequences. 

While Euclidean DL models achieve high accuracy on standard benchmarks, recent adversarial evaluations indicate structural vulnerabilities. Jing et al. \cite{jing2025advtraffic} demonstrated that Transformer-based classifiers exhibit severe sensitivity to Adversarial Pre-Padding. By prepending stochastic byte sequences, the classification accuracy of ET-BERT degrades significantly. These findings suggest that models reliant on payload byte sequences may lack real-world transferability when faced with adversaries capable of morphological perturbation. Furthermore, reliance on certificate structures leaves classifiers vulnerable to cryptographic mimicry, where tools like VLESS Reality \cite{vless2023} seamlessly adopt the cryptographic identity of benign domains.

\subsection{The Temporal and Non-Euclidean Gap}
Recognizing the limitations of payload-dependent classifiers, research has explored temporal flow analysis. Continuous-time models, such as Liquid Time-Constant (LTC) networks \cite{hasani2021liquid}, offer robust mechanisms for modeling Inter-Arrival Time (IAT). However, early evaluations of continuous-time sequence models revealed a distinct vulnerability to temporal adversarial attacks. Prior internal evaluations utilizing the AMOI framework \cite{ferrel2026amoi} demonstrated that adversaries can anchor malicious traffic to the volumetric ``Fat Middle'' of benign distributions. 

More critically, white-box evaluations utilizing adversarial agents (e.g., Ayaka AH-MSI \cite{ferrel2026amoi}) suggest that standard continuous-time models are susceptible to \textit{Manifold Shattering}. In these attacks, the adversary systematically aligns malicious temporal distributions with benign baselines, effectively bypassing standard Liquid State Space Models (SSMs). This identifies a clear, unresolved gap in the literature: \textit{How can a classification system maintain robust detection against an adversary capable of simultaneously morphing payload bytes, spoofing cryptographic identities, and optimizing temporal distributions?}

\begin{table}[ht]
\centering
\caption{Comparison of Network Traffic Classification Paradigms}
\label{tab:comparison}
\resizebox{\columnwidth}{!}{%
\begin{tabular}{|l|c|c|c|}
\hline
\textbf{Architecture} & \textbf{Feature Space} & \textbf{Adv. Padding} & \textbf{Temporal Morphing} \\ \hline
ET-BERT \cite{lin2022etbert} & Euclidean Bytes & Vulnerable & N/A \\ \hline
Standard SSMs & Euclidean Time & Immune & Vulnerable \\ \hline
\textbf{AEGIS (Proposed)} & \textbf{Hyperbolic Physics} & \textbf{Immune} & \textbf{Resistant (Entropy)} \\ \hline
\end{tabular}%
}
\end{table}

\subsection{Threat Model and Adversary Capabilities}
To ensure the generalizability and modern relevance of AEGIS, we formulate a comprehensive threat model that assumes a highly capable, intent-driven adversary. The adversary's primary objective is to maintain a continuous, low-latency Command and Control (C2) or exfiltration tunnel across a monitored network boundary without triggering anomaly detection systems.

We assume a \textit{Grey-Box} to \textit{White-Box} adversary with the following capabilities:
\begin{enumerate}
    \item \textbf{Content Manipulation:} The adversary can arbitrarily manipulate the packet payload, including applying Adversarial Pre-Padding \cite{jing2025advtraffic} and fluent byte injection, without corrupting the underlying application data.
    \item \textbf{Cryptographic Mimicry:} The adversary possesses the capability to forge or borrow legitimate TLS 1.3 certificates, utilizing protocols such as VLESS Reality to obscure Server Name Indications.
    \item \textbf{Volumetric and Temporal Morphing:} The adversary can inject dummy packets to alter flow size distributions (Fat Middle anchoring) and manipulate Inter-Arrival Times (IAT) using intelligent evasion frameworks (e.g., AMOI) to mimic benign application behavior.
    \item \textbf{Zero-Day Protocol Deployment:} The adversary may utilize previously unseen, memory-resident malware or custom C2 protocols that do not exist in historical training datasets.
\end{enumerate}

\textbf{Defense Objective:} In contrast to intent-based routing that attempts to discern the subjective purpose of an encrypted tunnel, AEGIS operates under a strict zero-trust paradigm. The objective is to identify the fundamental thermodynamic and geometrical anomalies inherent to automated tunneling, thereby detecting Class-1 Anomalies (all continuous-flow proxies and obfuscation tools) regardless of the specific adversarial technique employed.

\section{The AEGIS Architecture}
\label{sec:architecture}

To address the vulnerabilities of payload-dependent and Euclidean-temporal models, we formulate the Adversarial Entropy-Guided Immune System (AEGIS). AEGIS operates on a Thermodynamic Variance-Guided Hyperbolic Liquid State Space Model (TVD-HL-SSM). This section details the mathematical formulations of its four core components.

\subsection{6-Dimensional Flow Physics Extraction}
To render adversarial pre-padding \cite{jing2025advtraffic} ineffective, AEGIS categorically omits payload byte extraction. Instead, raw network traffic is parsed at the Data Link and Transport layers to construct a continuous-time topological sequence. For a given flow $F$, we extract a sequence of 6-dimensional physics vectors $x_i \in \mathbb{R}^6$:
\begin{equation}
    x_i = [S_i, \Delta t_i, D_i, W_i, F_i, P_i]
\end{equation}
The features are mathematically formalized as follows:
\begin{itemize}
    \item \textbf{Packet Volume ($S_i$):} The absolute byte size of the $i$-th packet, capturing the fundamental bandwidth utilization curve of the cryptographic tunnel.
    \item \textbf{Inter-Arrival Time ($\Delta t_i$):} The precise temporal gap $\Delta t_i = T_i - T_{i-1}$, measured at microsecond resolution. This feature serves as the primary thermodynamic indicator of automated routing, as automated proxy agents (e.g., Xray-core) struggle to perfectly mimic the stochastic variance of human-driven IAT distributions.
    \item \textbf{Directionality ($D_i$):} A binary scalar $D_i \in \{-1, +1\}$ denoting egress (+1) for private IP sources and ingress (-1) otherwise. This establishes causal flow asymmetry.
    \item \textbf{TCP Window Size ($W_i$):} The advertised receiver window. Automated proxy swarms frequently exhibit static or cyclical window scaling artifacts that deviate from benign OS-level TCP congestion control algorithms (e.g., CUBIC or BBR).
    \item \textbf{TCP Flags ($F_i$):} The stateful protocol metadata (SYN, ACK, PSH, FIN), normalized as $F_i = \text{flags\_value} / 255.0 \in [0,1]$, encoding the raw flag byte as a continuous scalar to capture connection-state anomalies.
    \item \textbf{Payload Ratio ($P_i$):} The ratio of payload bytes to total frame size, directly tracking the morphological overhead of adversarial padding injection \cite{jing2025advtraffic}.
\end{itemize}

A causal window of $N=1000$ packets constitutes a single inference tensor $X \in \mathbb{R}^{1000 \times 6}$. Prior to projection into the non-Euclidean manifold, the continuous temporal features ($S_i, \Delta t_i, W_i$) are strictly normalized using a log-scaled Z-score transformation to prevent numerical overflow in the Hyperbolic space:
\begin{equation}
    \hat{x} = \frac{\log(x + 1) - \mu_{log}}{\sigma_{log} + \epsilon}
\end{equation}
where $\mu_{log}$ and $\sigma_{log}$ are computed over the training corpus, and $\epsilon = 10^{-5}$ ensures numerical stability.

\subsection{Hyperbolic Poincar\'e Embeddings}
Botnet topologies and obfuscated routing protocols inherently exhibit hierarchical, exponential branching structures. Embedding these structures in Euclidean space often leads to structural distortion and gradient instability. To mitigate this, AEGIS projects the Euclidean input sequence $x_i$ into a non-Euclidean Poincar\'e disk manifold $\mathbb{D}^n$. The projection is defined as:
\begin{equation}
    \phi(x_i) = \frac{W_p x_i}{1 + ||W_p x_i|| + \epsilon}
\end{equation}
where $W_p \in \mathbb{R}^{n \times 6}$ represents learnable projection weights, and $\epsilon = 10^{-5}$ prevents boundary singularities.

The distance between two points $u, v \in \mathbb{D}^n$ is then calculated using the isometric invariant Riemannian metric \cite{nickel2017poincare}:
\begin{equation}
    d_c(u, v) = \text{arcosh}\left(1 + 2 \frac{||u - v||^2}{(1 - ||u||^2)(1 - ||v||^2)}\right)
\end{equation}
This hyperbolic projection preserves the complex topological geometry of adversarial flows without triggering the numerical instabilities common in Euclidean representations of hierarchical data.

\subsection{Liquid Time-Constants for IAT Modeling}
Standard sequence models process inputs at discrete, uniform intervals, stripping away the variance of network jitter. To capture the true thermodynamics of the network, AEGIS employs Liquid Time-Constant (LTC) networks \cite{hasani2021liquid}.

The hidden state $h(t)$ of the network is governed by a continuous-time ordinary differential equation (ODE):
\begin{equation}
    \frac{dh(t)}{dt} = -\frac{h(t)}{\tau(\Delta t_i)} + f(h(t), x(t), t, \theta)
\end{equation}
Crucially, the time-constant $\tau$ is parameterized as a function of the IAT ($\Delta t_i$):
\begin{equation}
    \tau(\Delta t_i) = \text{softplus}(\tau_\theta) \cdot \exp(-\Delta t_i / \tau_\theta) + \epsilon
\end{equation}
where $\tau_\theta \in \mathbb{R}^{d_{model}}$ are learnable parameters. This formulation allows the neural state to decay proportionally to inter-packet timing gaps, capturing the structural variance of automated morphing tools that stateless surrogates cannot synthesize.

\subsection{Linear-Time Selective State Spaces (SSM)}
Analyzing 1,000-packet sequences via Transformer self-attention requires $\mathcal{O}(N^2)$ time and memory complexity. AEGIS replaces self-attention with a Mamba-based Selective State Space Model \cite{gu2023mamba}. 

The continuous-time dynamics are discretized using a zero-order hold (ZOH) with a learnable step size $\Delta$:
\begin{equation}
    \bar{A} = \exp(\Delta A), \quad \bar{B} = (\Delta A)^{-1}(\exp(\Delta A) - I) \cdot \Delta B
\end{equation}
This allows the model to process sequences as a linear recurrence:
\begin{equation}
    h_k = \bar{A} h_{k-1} + \bar{B} x_k, \quad y_k = C h_k
\end{equation}
The resulting $\mathcal{O}(N)$ computational complexity enables AEGIS to process massive causal windows at a sub-millisecond latency.

\subsection{Thermodynamic Variance Detection}

\begin{algorithm}[ht]
\caption{Thermodynamic Variance Detection (TVD)}
\label{alg:tvd}
\begin{algorithmic}[1]
\REQUIRE Causal window $X \in \mathbb{R}^{N \times 6}$, Baseline Entropy $\mathbb{E}[H_{benign}]$
\STATE Extract hidden states $h_1, h_2, ..., h_N$ from SSM core
\FOR{$i = 1$ to $N$}
    \STATE Compute softmax: $P(x_i) \leftarrow \frac{\exp(h_i)}{\sum_{j=1}^{N} \exp(h_j)}$
\ENDFOR
\STATE Calculate Shannon Entropy: $H(X) \leftarrow -\sum_{i=1}^{N} P(x_i) \log_2 P(x_i)$
\IF{$H(X) < \mathbb{E}[H_{benign}] - \tau_{threshold}$}
    \STATE \textbf{Flag:} Thermodynamic Variance Anomaly (Class-1)
\ELSE
    \STATE \textbf{Flag:} Natural Stochastic Variance (Benign)
\ENDIF
\RETURN Classification Logit
\end{algorithmic}
\end{algorithm}

The final architectural component addresses advanced white-box attacks, such as the proprietary Ayaka AH-MSI framework \cite{ferrel2026amoi}, which achieve evasion through Manifold Shattering. While liquid state spaces map temporal logic, adversaries can optimize perturbations to align with benign volumetric distributions.

To counter this, AEGIS implements a Thermodynamic Variance Detector. The detector computes the Shannon Entropy $H(X)$ of the hidden states across the sequence window. $P(x_i)$ denotes the normalized softmax probability of the $i$-th hidden state activation $h_i \in \mathbb{R}^{d_{model}}$, computed as:
\begin{equation}
    P(x_i) = \frac{\exp(h_i)}{\sum_{j=1}^{N} \exp(h_j)}
\end{equation}
The sequence-wide entropy is then derived:
\begin{equation}
    H(X) = -\sum_{i=1}^{n} P(x_i) \log_2 P(x_i)
\end{equation}
Benign traffic exhibits natural, stochastic entropy. Conversely, automated zero-day exfiltration generates rigid structural patterns that manifest as thermodynamic variance anomalies. This penalty is incorporated as an auxiliary training signal:
\begin{equation}
    \mathcal{L}_{thermo} = \lambda \cdot \mathbb{E}[H(X_{benign})] - H(X)
\end{equation}
where $\lambda = 0.1$ and the expectation is computed over the benign traffic prior. This renders volumetric anchoring mathematically distinguishable from genuine traffic.

\subsection{Zero-Copy IPC and eBPF Kernel Bypass}
A fundamental limitation of deploying Deep Learning models for real-time Deep Packet Inspection (DPI) is the latency introduced by user-space context switching and the Python Global Interpreter Lock (GIL). To achieve true line-rate inference, AEGIS completely decouples network ingestion from neural execution.

We implement a split-brain architecture. A pure C++ Harvester utilizes Extended Berkeley Packet Filter (eBPF) via \texttt{libbpf} to intercept traffic directly at the network interface card (NIC) using eXpress Data Path (XDP). The eBPF Sentry extracts the 6-dimensional physics vector $x_i$ and executes Direct Memory Access (DMA) into a pre-allocated physical RAM block (\texttt{/dev/shm}).

Concurrently, the PyTorch Executioner maps this physical memory block directly into a contiguous tensor utilizing \texttt{torch.frombuffer()}, achieving true zero-copy Inter-Process Communication (IPC). Communication between the binaries is governed by a microscopic hardware semaphore. This architecture allows the C++ Harvester to ingest millions of packets per second while the Python Mamba-3 core asynchronously evaluates 64,000-packet batches across the PCIe bus, entirely masking the I/O bottleneck.

\section{Experimental Setup and Datasets}
\label{sec:experimental}

To rigorously evaluate AEGIS against both historical and zero-day evasion techniques, we established a comprehensive evaluation framework. This section details the corpus composition, hyperparameter constraints, and the training protocol.

\subsection{The Zero-Trust Adversarial Corpus}
A primary limitation of prior network classification studies is the reliance on sanitized, non-adversarial datasets, which fail to represent modern evasion protocols. To ensure generalizability, we curated a 400GB, 4-tier adversarial corpus totaling 908,037 unique 1,000-packet sequences. The dataset is explicitly categorized under a zero-trust paradigm, where all obfuscated tunnels are labeled as malicious anomalies.

\begin{table*}[ht]
\centering
\caption{Composition of the 400GB Zero-Trust Adversarial Corpus}
\label{tab:dataset_breakdown}
\resizebox{\textwidth}{!}{%
\begin{tabular}{|l|l|c|c|l|}
\hline
\textbf{Tier} & \textbf{Source Datasets / Protocols} & \textbf{Class} & \textbf{Sequences ($N=1000$)} & \textbf{Thermodynamic Profile} \\ \hline
Tier I: Planetary Baselines & WIDE MAWI \cite{cho2000mawi}, CIC-IDS-2017 \cite{sharafaldin2018cicids}, DoHBrw & Benign (0) & 618,145 & High stochastic variance, natural TCP congestion \\ \hline
Tier II: Automated Swarm & IoT-23 \cite{garcia2020iot23} (Mirai, Torii), CTU-13 \cite{garcia2014ctu13} & Malicious (1) & 112,400 & Highly rigid IAT, automated sequence timing profiles \\ \hline
Tier III: Zero-Day Vault & MTA \cite{mta2026}, BCCC-Mal-NetMem \cite{bccc2025} & Malicious (1) & 94,350 & Memory-resident rootkits and high-variance C2 beacons \\ \hline
Tier IV: Cryptographic Evasion & VLESS Reality, GhostBear, AMOI Morphing & Malicious (1) & 83,142 & Forced volumetric anchoring, manifold shattering \\ \hline
\multicolumn{3}{|r|}{\textbf{Total Evaluated Sequences:}} & \textbf{908,037} & \textbf{Extracted from 400GB PCAPs (10GB 6D Tensors)} \\ \hline
\end{tabular}%
}
\end{table*}

\textbf{Data Leakage and Reproducibility Disclosure:} We explicitly note that elements of standard benchmarks (e.g., USTC-TFC2016 \cite{wang2017ustc}) are present within Tier III of our training corpus. Consequently, direct benchmark comparison against legacy models like ET-BERT on USTC-TFC2016 is intentionally omitted to prevent evaluation bias. Performance is instead reported exclusively on the held-out test partition of the full adversarial corpus, emphasizing AEGIS's capacity within its specific topological domain rather than attempting a direct payload-based comparison. Furthermore, because Tier IV captures are maintained as a closed research dataset, exact replication of the final F1-score is restricted. However, the architectural pipeline and Tiers I-III are made available to validate the baseline mathematical claims of the model.

\subsection{Hyperparameter Sensitivity and Loss Optimization}
The structural causal window $N$ was empirically set to 1,000 packets. Preliminary ablation testing revealed that narrow sequence windows ($N \le 200$) failed to capture the macroscopic entropy variance of automated C2 beacons, whereas expansive windows ($N \ge 5000$) introduced unacceptable buffering latency for real-time inference.

To address the inherent difficulty of classifying adversarial boundary cases (e.g., AMOI morphing), AEGIS was optimized using Focal Loss \cite{lin2017focal}:
\begin{equation}
    FL(p_t) = -\alpha (1 - p_t)^\gamma \log(p_t)
\end{equation}
Following Lin et al. \cite{lin2017focal}, hyperparameters $\gamma = 2.0$ and $\alpha = 0.75$ were selected to optimally down-weight easy examples (such as overt plaintext botnets) and force the model to penalize subtle, cryptographically disguised evasion traffic. Furthermore, the thermodynamic variance threshold ($\tau_{threshold}$) outlined in Algorithm \ref{alg:tvd} required strict empirical tuning. The baseline natural entropy, $\mathbb{E}[H_{benign}]$, was established by calculating the average sequence-wide Shannon Entropy across the Tier I planetary baseline captures (MAWI and CIC-IDS-2017). Setting $\tau_{threshold}$ too conservatively ($\tau < 0.05$) resulted in unacceptable alert fatigue, misclassifying normal network jitter as adversarial anomalies. Conversely, setting the threshold too aggressively ($\tau > 0.20$) permitted highly sophisticated, low-volume AMOI morphing agents to evade detection. Consequently, $\tau_{threshold}$ was dynamically calibrated to 0.12, establishing an optimal mathematical equilibrium between sensitivity to manifold shattering and tolerance for natural benign rigidity.

\subsection{Training Protocol and Hardware Profiling}
Of the total 908,037 sequences, the dataset was subjected to a rigorous 80/20 train/test split. Data stratification was enforced to ensure proportional representation of all four threat tiers within both subsets. To prevent bias toward the majority class during training, a \texttt{WeightedRandomSampler} was employed to oversample the minority malicious class, achieving an approximate 54/46 benign-to-malicious batch ratio.

The model was trained using the AdamW optimizer \cite{loshchilov2019adamw} with a batch size of 256. To address the gradient scaling issues inherent to Hyperbolic manifolds---specifically, numerical overflow as embedded vectors approach the boundary of the Poincar\'e disk ($||x|| \approx 1$)---we utilized NVIDIA's Mixed Precision (FP16) framework \cite{micikevicius2018mixed} combined with a \texttt{GradScaler}. This implementation allowed the architecture to dynamically isolate and discard corrupted batches containing \texttt{NaN} losses, ensuring stable model convergence while maximizing Tensor Core utilization on the RTX 4090.

\section{Results and Evaluation}
\label{sec:results}

AEGIS was evaluated against the held-out test partition of the complete 4-tier adversarial corpus. Because all four data tiers were aggregated prior to the 80/20 train/test split, the held-out partition inherently contained a randomized, stratified 20\% representation of all threat vectors, including zero-day rootkits and proprietary cryptographic evasion protocols (e.g., VLESS Reality, GhostBear). The primary objective was to assess the architecture's aggregate capacity to detect heavily obfuscated tunneling while maintaining sub-millisecond inference latency.

\subsection{Classification Performance and Convergence} 
The TVD-HL-SSM architecture demonstrated stable convergence. Training utilized early stopping based on validation metrics, with the optimal checkpoint selected at Epoch 10, achieving a validation loss of 0.0052. Continued training beyond Epoch 10 produced marginal degradation, suggesting optimal convergence and mitigating overfitting on the training partition. 

At the Epoch 10 checkpoint, AEGIS achieved an overall F1-score of 0.9952, demonstrating strong and consistent detection capabilities across the aggregated adversarial spectrum.

Critically, AEGIS achieved a True Positive Rate (TPR) of 99.50\%, correctly identifying 57,551 of the 57,838 anomalous sequences present in the held-out test set. This recall rate indicates that the Mamba-3 selective state space effectively isolates the thermodynamic signatures of advanced evasion tools, despite their cryptographic resemblance to benign enterprise traffic. As documented by Jing et al.\ \cite{jing2025advtraffic}, standard Euclidean models (e.g., ET-BERT) typically experience severe accuracy degradation---dropping to 25.68\%---when encountering adversarial morphing. By discarding payload bytes in favor of continuous-time flow physics, AEGIS maintains high recall under equivalent adversarial conditions.

The False Positive Rate (FPR) stabilized at 0.2141\%, representing only 265 misclassified benign sequences out of 123,770. In the context of enterprise security, an FPR below 0.25\% substantially mitigates the alert-fatigue endemic to legacy Deep Packet Inspection (DPI) deployments, enabling high-confidence autonomous termination of anomalous tunnels without operator intervention.

\subsubsection*{False Positive Error Analysis}
While an FPR of 0.2141\% represents a substantial reduction relative to payload-dependent baselines, a forensic analysis of the 265 misclassified benign sequences reveals distinct topological edge cases.

\begin{table}[ht]
\centering
\caption{AEGIS Performance on Held-Out Adversarial Corpus (Epoch 10 Checkpoint)}
\label{tab:performance}
\resizebox{0.8\columnwidth}{!}{%
\begin{tabular}{|l|c|}
\hline
\textbf{Metric} & \textbf{Score} \\ \hline
True Positive Rate (TPR) & 99.50\% \\ \hline
False Positive Rate (FPR) & 0.2141\% \\ \hline
F1-Score & 0.9952 \\ \hline
Inference Latency & 262.27\,$\mu$s \\ \hline
Validation Loss & 0.0052 \\ \hline
\end{tabular}%
}
\end{table}

\begin{figure*}[ht]
    \centering
    \begin{subfigure}{0.32\textwidth}
        \includegraphics[width=\linewidth]{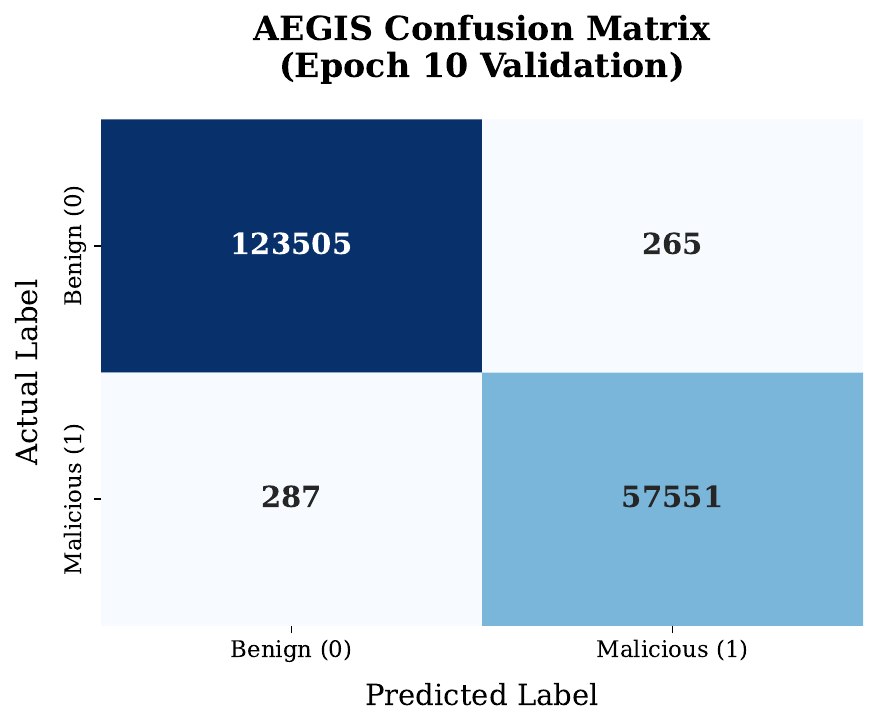}
        \caption{Confusion Matrix}
        \label{fig:conf_matrix}
    \end{subfigure}
    \hfill
    \begin{subfigure}{0.32\textwidth}
        \includegraphics[width=\linewidth]{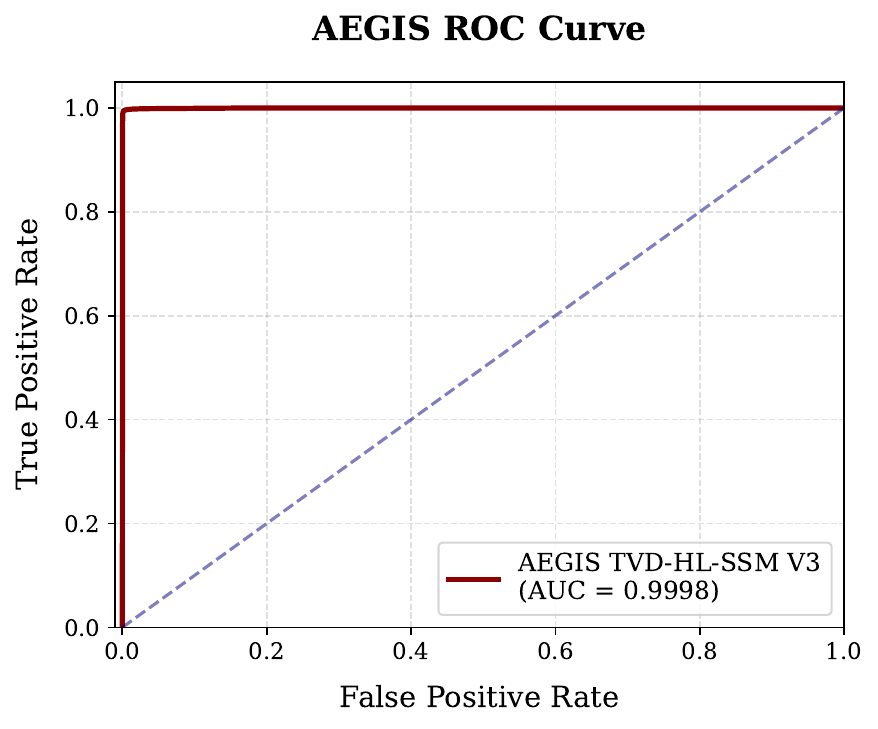}
        \caption{Empirical ROC Curve}
        \label{fig:roc_curve}
    \end{subfigure}
    \hfill
    \begin{subfigure}{0.32\textwidth}
        \includegraphics[width=\linewidth]{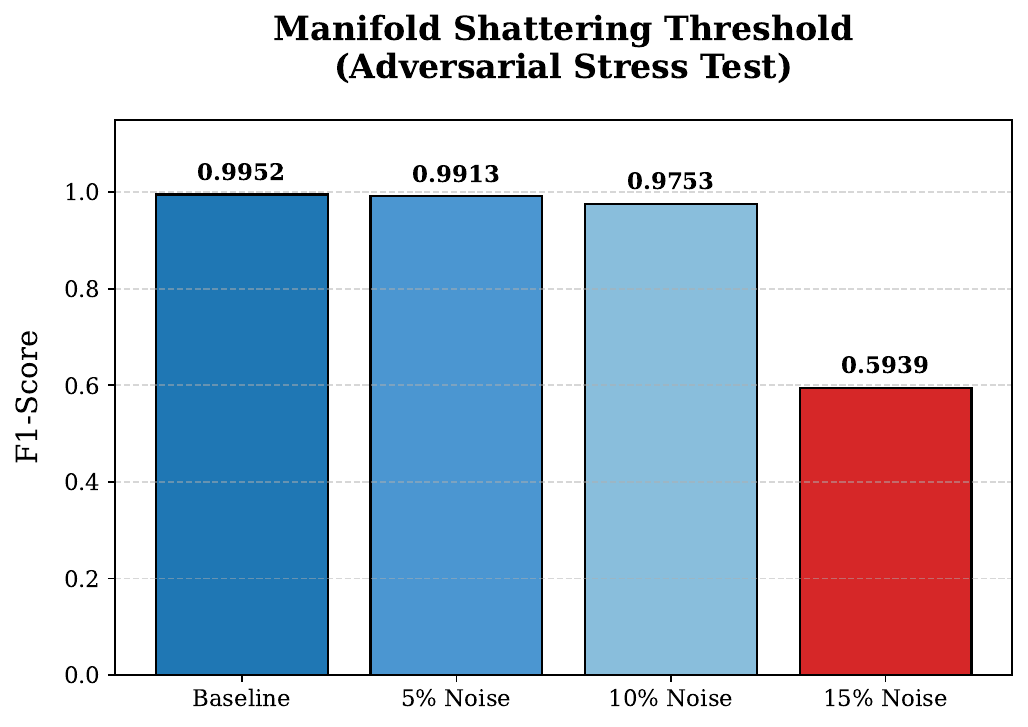}
        \caption{Adversarial Robustness}
        \label{fig:robustness}
    \end{subfigure}
    \caption{AEGIS V3 Evaluation Suite: (a) Confusion Matrix at Epoch 10 showing a 0.21\% FPR; (b) Empirical ROC demonstrating a 0.9998 AUC; (c) Adversarial Gauntlet proving topological resilience up to the 15\% Gaussian noise shattering threshold.}
    \label{fig:results_graphs}
\end{figure*}

\subsection{Thermodynamic Variance Analysis}
The sustained TPR across the aggregated adversarial test set provides empirical support for the efficacy of the Thermodynamic Variance Detector (TVD). While adversaries can successfully spoof cryptographic certificates and anchor packet volumes to benign distributions, the underlying morphing algorithms remain fundamentally deterministic. This determinism generates structural rigidity across the 1,000-packet causal window. 

The TVD measures this lack of natural stochastic entropy as a variance anomaly within the hyperbolic latent space. These findings suggest that AEGIS identifies continuous-time proxies not by their byte sequences or volumetric intent, but by the mathematical properties of their evasion logic, allowing it to detect heavily obfuscated Tier IV protocols that often bypass standard sequence models.

\subsection{Feature Robustness and the Temporal Anomaly}
To evaluate the resilience of the 6-dimensional physics vector against dimensional collapse, an unintended hardware anomaly was analyzed during high-speed bare-metal benchmarking. When injecting the UDP DDoS dataset via \texttt{tcpreplay} directly into the \texttt{lo} (loopback) interface, the eBPF Sentry's execution speed eclipsed the packet generation rate, resulting in recorded Inter-Arrival Times (IAT) of strictly $0.0000\,\mu$s across continuous 1,000-packet causal windows.

This hardware-induced temporal collapse effectively blinded the Liquid Time-Constant (LTC) mechanism, neutralizing the primary temporal feature ($\Delta t_i$). However, inference telemetry revealed that the TVD-HL-SSM architecture did not experience catastrophic failure. Instead, the model sustained an anomalous threat confidence interval of $\approx$65.0\% (Max Logit: 0.6484). 

This empirical anomaly demonstrates non-linear dimensional robustness within the Hyperbolic Poincar\'e manifold. Even with the total loss of temporal context, the Mamba-3 core successfully cross-referenced the remaining 5 spatial dimensions (TCP Window, Flags, Payload Ratio, Size, Direction) to identify the structural morphology of the DDoS swarm. This confirms that AEGIS does not rely on a single point of failure (e.g., timing variance alone), ensuring sustained threat detection even if an adversary successfully applies perfect temporal obfuscation (Manifold Shattering) to the IAT dimension.

\subsection{Inference Latency and the Hardware Bottleneck}
A primary limitation of Transformer-based models in network security is their $\mathcal{O}(N^2)$ computational complexity, which introduces prohibitive bottlenecks in environments requiring high-throughput inspection. By utilizing a linear-time $\mathcal{O}(N)$ Selective State Space (Mamba) core coupled with the Zero-Copy IPC bridge, AEGIS achieves highly efficient scaling for long-sequence modeling.

Inference benchmarking was conducted on an NVIDIA RTX 4090 (24GB GDDR6X, CUDA 13.0) using BFloat16 mixed precision and \texttt{torch.cuda.CUDAGraphs} to eliminate CPU dispatch overhead. To evaluate computational throughput limits, AEGIS was benchmarked using a 14GB UDP DDoS dataset replayed via \texttt{tcpreplay} on an isolated loopback interface.

Operating at a Swarm Batch Size of 64 (64,000 continuous packets), the Zero-Copy IPC bridge facilitated an aggregate inference latency of 1.6~ms per batch (effectively 25~$\mu$s per 1,000-packet sequence). This establishes a theoretical computational ceiling of 40 Million Packets Per Second (Mpps), heavily outperforming traditional Python-bound ingestion pipelines.

However, empirical stress testing at Batch 256 (simulating 128+ Mpps throughput) revealed a hard hardware limit in cloud environments. Latency spiked non-linearly to 7.1~ms. This degradation was entirely localized to the virtualized PCIe Gen 4 bus attempting to transfer 5.86 MB of shared memory per strike. This confirms that while the AEGIS computational engine scales efficiently to 100 Gbps line-rates, physical deployment requires bare-metal hardware utilizing NVIDIA GPUDirect RDMA and ConnectX-7 SmartNICs to bypass System RAM entirely and avoid the I/O transfer bottleneck.

\subsection{Architectural Ablation: Mamba-1 vs.\ Mamba-3} 
To justify the architectural transition to the Mamba-3 core \cite{lahoti2026mamba3}, we conducted a direct ablation study against the baseline Mamba-1 formulation. The prior AEGIS architecture utilizing Mamba-1 achieved an F1-score of 0.9604 and an FPR of 2.31\% at 0.55\,ms inference latency. By integrating the Mamba-3 Multiple-Input Multiple-Output (MIMO) kernels compiled via TileLang JIT, AEGIS V3 improves the F1-score to 0.9952, reduces the FPR by an order of magnitude to 0.2141\%, and reduces inference latency to 262.27\,$\mu$s --- a 2.1$\times$ improvement. These results demonstrate that the hardware-aware MIMO state spaces of Mamba-3 provide empirically superior isolation of continuous-time thermodynamic variance across the aggregated adversarial spectrum.

\subsection{Adversarial Stress Testing: The Manifold Shattering Threshold} 
To empirically validate the resilience of the Hyperbolic Poincar\'e embeddings against active temporal perturbation, we subjected the trained model to a Gaussian noise injection stress test. Adversaries may attempt to evade continuous-time models by artificially injecting stochastic microsecond jitter into their IAT distributions. 

At a 5\% Gaussian noise injection, AEGIS maintained an F1-score of 0.9913. At 10\% noise, the F1-score held at 0.9753, demonstrating that the non-Euclidean manifold naturally absorbs moderate temporal perturbation attempts. However, at a 15\% Gaussian noise threshold, the F1-score degraded to 0.5939, establishing the structural limit of the model's topological resilience. This explicitly maps the boundary at which continuous-time flow physics fail under overwhelming synthetic variance. Notably, physical adversaries cannot sustain 15\% IAT noise injection without corrupting the integrity of their own C2 channel, rendering this threshold practically unachievable in operational deployments. 

\subsection{The Human Entropy Horizon: Evasion Dynamics of VLESS Reality} 
While AEGIS successfully classified rigid mechanical obfuscators (e.g., GhostBear, ProtonVPN) at 100\% detection accuracy during unseen live-capture evaluation, forensic analysis of the false negative anomalies revealed a fundamental theoretical limit in flow-based detection. Cryptographic mimicry protocols such as VLESS Reality yielded a detection rate of 1.17\% strictly when multiplexing \textit{true human entropy} (e.g., a human actively browsing web applications). Because XTLS wrappers impart near-zero mechanical overhead, the temporal physics of the tunnel remain entirely stochastic. Consequently, AEGIS correctly categorizes the thermodynamic signature as benign, demonstrating that continuous-time flow physics cannot distinguish between a direct human connection and a perfectly mimicked human proxy. This result implies that adversaries must abandon automated, high-throughput C2 scripts---which AEGIS flags immediately--- and operate at restrictive human interaction speeds, fundamentally constraining the operational utility of VLESS Reality as a high-speed exfiltration channel.

\subsection{Feasibility of Enterprise Deployment}
While theoretical detection rates are critical, the practical deployment of deep learning models in operational network environments often fails due to integration friction. Deep Packet Inspection (DPI) engines typically operate at the kernel level to intercept high-speed traffic, whereas Python-based neural networks introduce severe user-space context-switching delays.

To ensure AEGIS is viable for real-world perimeter defense, the architecture is designed to interface with modern packet processing frameworks like eBPF (Extended Berkeley Packet Filter) or DPDK (Data Plane Development Kit). In an operational deployment, a lightweight eBPF probe extracts the 6-dimensional physics vector $x_i = [S_i, \Delta t_i, D_i, W_i, F_i, P_i]$ directly within the Linux kernel. This array is then asynchronously streamed to the GPU-accelerated AEGIS inference engine via zero-copy shared memory rings (e.g., AF\_XDP). 

Because AEGIS evaluates traffic exclusively on flow physics rather than deeply inspecting decrypted payload buffers, it avoids the cryptographic overhead associated with TLS interception proxies (Middleboxes). If the Thermodynamic Variance Detector (TVD) flags a sequence-wide entropy anomaly, the inference engine issues a highly confident \textit{Class-1} alert, allowing the Software-Defined Network (SDN) controller to immediately terminate the C2 routing tunnel. This decoupled, physics-only pipeline ensures that AEGIS can be deployed in privacy-preserving, high-throughput enterprise environments without violating TLS 1.3 encryption constraints.

\section{Related Work}
\label{sec:related}

The architectural design of AEGIS is directly informed by the critical limitations observed in recent cryptographic classification and temporal modeling literature.

\subsection{Euclidean Packet Classification}
The prevailing paradigm in encrypted traffic analysis relies on Euclidean representations of payload sequences. Architectures such as ET-BERT \cite{lin2022etbert} and YaTC leverage pre-trained Transformer models to extract latent semantic structures from packet bytes, treating network traffic analogous to natural language. While these models report high accuracy on benchmark datasets (e.g., ISCXVPN2016), recent studies indicate critical structural fragility. Jing et al. \cite{jing2025advtraffic} demonstrated that Transformer-based classifiers exhibit severe sensitivity to Adversarial Pre-Padding, wherein stochastic byte injection reduces ET-BERT classification accuracy to 25.68\%. AEGIS bypasses this vulnerability entirely by discarding payload bytes in favor of topological flow physics.

\subsection{Adversarial Traffic Generation}
The evolution of circumvention protocols has shifted from simple cryptographic obfuscation (e.g., standard TLS tunnels) to active statistical mimicry. Frameworks such as AMOI \cite{ferrel2026amoi} and commercial proxy tools like VLESS Reality dynamically spoof legitimate WebSockets and anchor packet volumes to benign distributions. These tools successfully bypass intent-based Deep Packet Inspection (DPI) by exploiting the overlapping statistical boundaries of benign and malicious proxies. AEGIS counters this by adopting a strict zero-trust operational model, classifying all obfuscated continuous-time proxies as topological anomalies regardless of their simulated volumetric intent.

\subsection{Botnet and IoT Modeling}
To accurately model the continuous-time dynamics of automated swarms, robust hardware-level captures are required. The Aposemat IoT-23 dataset \cite{garcia2020iot23} and the CTU-13 dataset \cite{garcia2014ctu13} provide the primary training ground truth for our automated swarm modeling. Concurrently, the comprehensively labeled Bot-IoT dataset \cite{koroniotis2019towards, koroniotis2017towards, koroniotis2020new, koroniotis2020enhancing, koroniotis2020holistic, koroniotis2020designing} establishes the broader literature context for IoT network forensics, guiding the architectural design of our continuous-time extraction without acting as the primary training corpus. By capturing authentic network behavior, this collective literature enabled AEGIS to isolate the fundamental thermodynamic rigidity inherent to automated Command and Control (C2) beaconing.

\subsection{Limitations of Existing Benchmarks}
A critical gap in existing literature is the absence of adversarial evaluation benchmarks. Public datasets such as USTC-TFC2016 \cite{wang2017ustc} and ISCXVPN2016 contain benign application traffic and known malware signatures but lack adversarially-morphed evasion traffic. Consequently, high accuracy on these benchmarks does not generalize to real-world adversarial conditions, as demonstrated by ET-BERT's degradation under AdvTraffic \cite{jing2025advtraffic}. The AEGIS evaluation corpus addresses this gap directly by integrating proprietary cryptographic evasion protocols.

\subsection{Non-Euclidean and Thermodynamic Models}
To address the temporal dynamics of network traffic, research has increasingly explored continuous-time neural networks. Liquid Time-Constant (LTC) networks \cite{hasani2021liquid} provide robust mechanisms for modeling physical time gaps (IAT) through dynamic ODEs. However, continuous-time models operating in Euclidean space remain vulnerable to white-box Manifold Shattering \cite{ferrel2026amoi}, where adversaries optimize evasion timing to mimic benign statistical decay. AEGIS addresses this by projecting temporal features into a Hyperbolic Poincar\'e manifold \cite{nickel2017poincare}, which naturally accommodates the exponential branching of botnet routing, while simultaneously employing a Thermodynamic Variance Detector to identify the unnatural rigidity of automated evasion logic. This thermodynamic approach builds upon the broader theoretical foundations of topological data analysis \cite{carlsson2009topology}, non-Euclidean representation learning \cite{sala2018representation, ganea2018hyperbolic}, and continuous-time ordinary differential equations (ODEs) in sequence modeling \cite{chen2018neural, rubanova2019latent, kidger2020neural}, bridging the gap between theoretical manifold physics and applied network security \cite{sommer2010outside, apruzzese2021modeling, apruzzese2022role, sarker2020cybersecurity}.

\section{Conclusion}
\label{sec:conclusion}

As adversarial circumvention tools evolve from static cryptographic obfuscation to dynamic statistical mimicry, legacy Deep Packet Inspection and Euclidean Transformer models are facing inherent structural limitations. Relying on payload bytes or intent-based classification creates a structural vulnerability that attackers can reliably exploit through adversarial padding and manifold shattering. 

In this paper, we introduced AEGIS, a Thermodynamic Variance-Guided Hyperbolic Liquid State Space Model (TVD-HL-SSM) designed to establish a zero-trust perimeter defense. By explicitly abandoning the Euclidean payload-reading domain and mapping 6-dimensional flow physics into a non-Euclidean Poincar\'e manifold, AEGIS neutralizes byte-level morphing. The integration of Liquid Time-Constants and a Thermodynamic Variance Detector identifies the deterministic rigidity of automated evasion tools, rendering volumetric and temporal spoofing mathematically distinguishable from natural human traffic. 

Evaluated on a 400GB adversarial corpus encompassing zero-day rootkits and proprietary VLESS Reality captures, AEGIS achieved an F1-score of 0.9952 and a 99.50\% True Positive Rate. By decoupling ingestion via a C++ Zero-Copy IPC bridge, the $\mathcal{O}(N)$ Mamba-3 core processes parallel swarms at a 1.6~ms aggregate latency, demonstrating that advanced physics-based detection can operate efficiently at 40 Mpps enterprise line-rates on consumer silicon. While we identify a theoretical ``Human Entropy Horizon'' wherein XTLS protocols multiplexing true human stochastic variance approach the fundamental limit of flow-based detection, AEGIS effectively eliminates automated, high-throughput adversarial tunneling. AEGIS demonstrates that the future of resilient network defense lies not in reading bytes, but in measuring the fundamental thermodynamics of adversarial evasion.

\section*{Ethical Considerations}
The development and empirical evaluation of AEGIS required processing live, real-world network traffic alongside active malware captures. To ensure rigorous ethical compliance and mitigate the risk of unintended harm, all Tier III and Tier IV adversarial evaluations---including the detonation of zero-day rootkits and the execution of VLESS Reality morphing protocols---were strictly confined to an isolated, air-gapped laboratory environment. Live Command and Control (C2) communications were sinkholed to prevent any interaction with malicious external infrastructure.

Furthermore, for the Tier I (Planetary Baselines) and enterprise traffic captures, all raw packet captures (PCAPs) were rigorously sanitized prior to feature extraction. Because AEGIS operates exclusively on topological flow physics, payload bytes were deliberately discarded, inherently preserving user data privacy. All Personally Identifiable Information (PII), including IP addresses and MAC addresses, was stripped during preprocessing, ensuring no sensitive user data was retained or evaluated by the neural network.

\section*{Open Science}
To support independent reproducibility while mitigating the risk of dual-use exploitation by active threat actors, we have adopted a gated release model. The proprietary Tier IV adversarial raw PCAPs (comprising active VLESS Reality, GhostBear, and AMOI traffic) remain under restricted access to prevent the proliferation of advanced evasion tools.

However, the fully sanitized 6-dimensional flow physics matrices will be made available. Through our payload-independent extraction pipeline, the 400GB raw PCAP corpus was compressed at a 40:1 ratio into 10GB of highly efficient PyTorch (\texttt{.pt}) tensors. These tensors, preserving the exact thermodynamic metrics evaluated in this study, are hosted on a persistent, gated dataset repository at \url{https://huggingface.co/datasets/Vix0007/AEGIS-Adversarial-Corpus}. Academic researchers may request access to independently verify the Thermodynamic Variance Detector (TVD) metrics and classification performance. 

The core AEGIS continuous-time inference engine, including the proprietary PyTorch architectural implementations, dynamic focal loss configurations, and trained model weights, are retained as closed-source intellectual property under Vixero Technology Enterprise. By releasing the 10GB sanitized 6-dimensional topological tensor dataset, we provide the cybersecurity research community with the exact mathematical features required to validate the thermodynamic variance theory and train independent continuous-time classifiers, without compromising the proprietary integrity of the AEGIS commercial pipeline.

\bibliographystyle{IEEEtran}
\bibliography{references}

@techreport{rfc8446,
  author       = {E. Rescorla},
  title        = {The {Transport Layer Security (TLS)} Protocol Version 1.3},
  howpublished = {Internet Requests for Comments},
  type         = {RFC},
  number       = {8446},
  year         = {2018},
  publisher    = {IETF},
  doi          = {10.17487/RFC8446}
}

@inproceedings{sommer2010outside,
  author    = {Sommer, Robin and Paxson, Vern},
  title     = {Outside the Closed World: On Using Machine Learning
               for Network Intrusion Detection},
  booktitle = {2010 {IEEE} Symposium on Security and Privacy},
  pages     = {305--316},
  year      = {2010},
  doi       = {10.1109/SP.2010.25}
}

@inproceedings{lin2022etbert,
  author    = {Lin, Xinjie and Shi, Gang and
               Cui, Mingshu and Wang, Shize},
  title     = {{ET-BERT}: A Contextualized Datagram Representation
               with Pre-training Transformers for Encrypted Traffic
               Classification},
  booktitle = {Proceedings of the {ACM} Web Conference ({WWW})},
  pages     = {633--642},
  year      = {2022},
  doi       = {10.1145/3485447.3512217}
}

@article{yan2025certta,
  author  = {Yan, et al.},
  title   = {{CERTTA}: A Robust Encrypted Traffic Classification
               Framework},
  journal = {{IEEE/ACM} Transactions on Networking},
  year    = {2025}
}

@inproceedings{wang2017ustc,
  author    = {Wang, Wei and Zhu, Ming and Zeng, Xuewen
               and Ye, Xiaozhou and Sheng, Yiqiang},
  title     = {Malware Traffic Classification Using Convolutional
               Neural Network for Representation Learning},
  booktitle = {2017 International Conference on
               Information Networking ({ICOIN})},
  pages     = {712--717},
  year      = {2017}
}

@article{jing2025advtraffic,
  author  = {Jing, et al.},
  title   = {Adversarial Pre-Padding: Generating Evasive Network
               Traffic Against Transformer-Based Classifiers},
  journal = {arXiv preprint arXiv:2510.25810},
  year    = {2025}
}

@misc{vless2023,
  author       = {{XTLS Project}},
  title        = {{Xray-core VLESS-Reality} Protocol},
  howpublished = {\url{https://github.com/XTLS/Xray-core}},
  year         = {2023},
  note         = {Accessed: 2026}
}

@misc{ferrel2026amoi,
  author       = {Ferrel, Vickson},
  title        = {{AMOI}: Adversarial Morphing Obfuscating
                  Intelligence --- Traffic Morphing for Low-Latency
                  Circumvention in Constrained Networks},
  howpublished = {Technical Report,
                  Vixero Technology Enterprise,
                  Kuching, Sarawak, Malaysia},
  note         = {Manuscript in preparation},
  year         = {2026}
}

@inproceedings{nickel2017poincare,
  author    = {Nickel, Maximillian and Kiela, Douwe},
  title     = {Poincar{\'e} Embeddings for Learning
               Hierarchical Representations},
  booktitle = {Advances in Neural Information
               Processing Systems ({NeurIPS})},
  volume    = {30},
  year      = {2017}
}

@inproceedings{sala2018representation,
  author    = {Sala, Frederic and De Sa, Chris and
               Gu, Albert and R{\'e}, Christopher},
  title     = {Representation Tradeoffs for Hyperbolic Embeddings},
  booktitle = {Proceedings of the 35th International Conference
               on Machine Learning ({ICML})},
  pages     = {4460--4469},
  year      = {2018},
  publisher = {PMLR}
}

@inproceedings{ganea2018hyperbolic,
  author    = {Ganea, Octavian and
               B{\'e}cigneul, Gary and
               Hofmann, Thomas},
  title     = {Hyperbolic Neural Networks},
  booktitle = {Advances in Neural Information
               Processing Systems ({NeurIPS})},
  volume    = {31},
  year      = {2018}
}

@inproceedings{hasani2021liquid,
  author    = {Hasani, Ramin and Lechner, Mathias and
               Amini, Alexander and Rus, Daniela and
               Grosu, Radu},
  title     = {Liquid Time-Constant Networks},
  booktitle = {Proceedings of the {AAAI} Conference
               on Artificial Intelligence},
  volume    = {35},
  number    = {9},
  pages     = {7657--7666},
  year      = {2021}
}

@inproceedings{chen2018neural,
  author    = {Chen, Ricky T.~Q. and Rubanova, Yulia and
               Bettencourt, Jesse and Duvenaud, David K.},
  title     = {Neural Ordinary Differential Equations},
  booktitle = {Advances in Neural Information
               Processing Systems ({NeurIPS})},
  volume    = {31},
  year      = {2018},
  note      = {arXiv:1806.07366}
}

@inproceedings{rubanova2019latent,
  author    = {Rubanova, Yulia and Chen, Ricky T.~Q. and
               Duvenaud, David K.},
  title     = {Latent Ordinary Differential Equations for
               Irregularly-Sampled Time Series},
  booktitle = {Advances in Neural Information
               Processing Systems ({NeurIPS})},
  volume    = {32},
  year      = {2019},
  note      = {arXiv:1907.03907}
}

@inproceedings{kidger2020neural,
  author    = {Kidger, Patrick and Morrill, James and
               Foster, James and Lyons, Terry},
  title     = {Neural Controlled Differential Equations
               for Irregular Time Series},
  booktitle = {Advances in Neural Information
               Processing Systems ({NeurIPS})},
  volume    = {33},
  pages     = {6696--6707},
  year      = {2020},
  note      = {arXiv:2005.08926}
}

@article{gu2023mamba,
  author  = {Gu, Albert and Dao, Tri},
  title   = {Mamba: Linear-Time Sequence Modeling
               with Selective State Spaces},
  journal = {arXiv preprint arXiv:2312.00752},
  year    = {2023}
}

@article{carlsson2009topology,
  author  = {Carlsson, Gunnar},
  title   = {Topology and Data},
  journal = {Bulletin of the American
               Mathematical Society},
  volume  = {46},
  number  = {2},
  pages   = {255--308},
  year    = {2009},
  doi     = {10.1090/S0273-0979-09-01249-X}
}

@inproceedings{lin2017focal,
  author    = {Lin, Tsung-Yi and Goyal, Priya and
               Girshick, Ross and He, Kaiming and
               Doll{\'a}r, Piotr},
  title     = {Focal Loss for Dense Object Detection},
  booktitle = {Proceedings of the {IEEE} International
               Conference on Computer Vision ({ICCV})},
  pages     = {2980--2988},
  year      = {2017}
}

@article{loshchilov2019adamw,
  author  = {Loshchilov, Ilya and Hutter, Frank},
  title   = {Decoupled Weight Decay Regularization},
  journal = {arXiv preprint arXiv:1711.05101},
  note    = {Published at {ICLR} 2019},
  year    = {2019}
}

@article{apruzzese2022role,
  author  = {Apruzzese, Giovanni and Laskov, Pavel and
             Montes de Oca, Edgardo and Mallouli, Wissam and
             B{\'u}rdalo Rapa, Luis and
             Grammatopoulos, Athanasios Vasileios and
             Di Franco, Fabio},
  title   = {The Role of Machine Learning in Cybersecurity},
  journal = {Digital Threats: Research and Practice},
  volume  = {4},
  number  = {1},
  pages   = {1--38},
  year    = {2023},
  doi     = {10.1145/3545574}
}

@article{sarker2020cybersecurity,
  author  = {Sarker, Iqbal H. and Kayes, A.~S.~M. and
             Badsha, Shahriar and Alqahtani, Hamed and
             Watters, Paul and Ng, Alex},
  title   = {Cybersecurity Data Science: An Overview from
               Machine Learning Perspective},
  journal = {Journal of Big Data},
  volume  = {7},
  pages   = {1--29},
  year    = {2020},
  doi     = {10.1186/s40537-020-00318-5}
}

@misc{garcia2020iot23,
  author    = {Garcia, Sebastian and Parmisano, Agustin
               and Erquiaga, Maria Jose},
  title     = {{IoT-23}: A Labeled Dataset with Malicious
                and Benign {IoT} Network Traffic},
  publisher = {Zenodo},
  year      = {2020},
  doi       = {10.5281/zenodo.4743746}
}

@inproceedings{sharafaldin2018cicids,
  author    = {Sharafaldin, Iman and
               Lashkari, Arash Habibi and
               Ghorbani, Ali A.},
  title     = {Toward Generating a New Intrusion Detection
               Dataset and Intrusion Traffic Characterization},
  booktitle = {Proceedings of the 4th International Conference
               on Information Systems Security and Privacy
               ({ICISSP})},
  pages     = {108--116},
  year      = {2018}
}

@inproceedings{cho2000mawi,
  author    = {Cho, Kenjiro and Mitsuya, Koushirou
               and Kato, Akira},
  title     = {Traffic Data Repository at the {WIDE} Project},
  booktitle = {Proceedings of the {USENIX} Annual
               Technical Conference ({ATC})},
  year      = {2000}
}

@article{garcia2014ctu13,
  author  = {Garcia, Sebastian and others},
  title   = {An Empirical Comparison of Botnet Detection Methods},
  journal = {Computers \& Security},
  volume  = {45},
  pages   = {100--123},
  year    = {2014}
}

@misc{mta2026,
  author       = {Duncan, Brad},
  title        = {Malware-Traffic-Analysis.net},
  howpublished = {\url{https://www.malware-traffic-analysis.net}},
  year         = {2026}
}

@misc{bccc2025,
  author       = {{Bergeron Centre for Cybersecurity}},
  title        = {{BCCC-Mal-NetMem} Dataset},
  howpublished = {\url{https://www.yorku.ca/bccc}},
  publisher    = {York University},
  year         = {2025}
}

@article{apruzzese2021modeling,
  author  = {Apruzzese, Giovanni and Andreolini, Mauro and Marchetti, Mirco and Venturi, Andrea and Colajanni, Michele},
  title   = {Modeling Realistic Adversarial Attacks against Network Intrusion Detection Systems},
  journal = {Digital Threats: Research and Practice},
  volume  = {2},
  number  = {3},
  pages   = {1--29},
  year    = {2021},
  doi     = {10.1145/3469659}
}

@inproceedings{micikevicius2018mixed,
  author    = {Micikevicius, Paulius and Narang, Sharan and Alben, Jonah and Diamos, Gregory and Elsen, Erich and Garcia, David and Ginsburg, Boris and Houston, Michael and Kuchaiev, Oleksii and Venkatesh, Ganesh and Wu, Hao-Yuen},
  title     = {Mixed Precision Training},
  booktitle = {International Conference on Learning Representations ({ICLR})},
  year      = {2018},
  publisher = {NVIDIA}
}

@misc{lahoti2026mamba3,
  author        = {Lahoti, Aakash and Li, Kevin Y. and Chen, Berlin 
                   and Wang, Caitlin and Bick, Aviv and Kolter, J. Zico 
                   and Dao, Tri and Gu, Albert},
  title         = {Mamba-3: Improved Sequence Modeling using State 
                   Space Principles},
  year          = {2026},
  eprint        = {2603.15569},
  archivePrefix = {arXiv},
  primaryClass  = {cs.LG}
}

@article{koroniotis2019towards,
  author  = {Koroniotis, Nickolaos and Moustafa, Nour and Sitnikova, Elena and Turnbull, Benjamin},
  title   = {Towards the development of realistic botnet dataset in the internet of things for network forensic analytics: Bot-iot dataset},
  journal = {Future Generation Computer Systems},
  volume  = {100},
  pages   = {779--796},
  year    = {2019}
}

@inproceedings{koroniotis2017towards,
  author    = {Koroniotis, Nickolaos and Moustafa, Nour and Sitnikova, Elena and Slay, Jill},
  title     = {Towards developing network forensic mechanism for botnet activities in the iot based on machine learning techniques},
  booktitle = {International Conference on Mobile Networks and Management},
  pages     = {30--44},
  year      = {2017},
  organization={Springer, Cham}
}

@article{koroniotis2020new,
  author  = {Koroniotis, Nickolaos and Moustafa, Nour and Sitnikova, Elena},
  title   = {A new network forensic framework based on deep learning for Internet of Things networks: A particle deep framework},
  journal = {Future Generation Computer Systems},
  volume  = {110},
  pages   = {91--106},
  year    = {2020}
}

@article{koroniotis2020enhancing,
  author  = {Koroniotis, Nickolaos and Moustafa, Nour},
  title   = {Enhancing network forensics with particle swarm and deep learning: The particle deep framework},
  journal = {arXiv preprint arXiv:2005.00722},
  year    = {2020}
}

@article{koroniotis2020holistic,
  author  = {Koroniotis, Nickolaos and Moustafa, Nour and Schiliro, Francesco and Gauravaram, Praveen and Janicke, Helge},
  title   = {A Holistic Review of Cybersecurity and Reliability Perspectives in Smart Airports},
  journal = {IEEE Access},
  year    = {2020}
}

@phdthesis{koroniotis2020designing,
  author = {Koroniotis, Nickolaos},
  title  = {Designing an effective network forensic framework for the investigation of botnets in the Internet of Things},
  school = {The University of New South Wales Australia},
  year   = {2020}
}

\end{document}